\documentclass[aps,prb,twocolumn,letterpaper,superscriptaddress,showpacs]{revtex4}
\usepackage{graphicx}
\usepackage{CJK}
\usepackage{amsmath,amssymb}

\usepackage{tikz}

\newcommand{\rlfao}{RbLn$_{2}$Fe$_{4}$As$_{4}$O$_{2}$ (Ln = Tb, Dy and Ho)}
\newcommand{\rgfao}{RbGd$_{2}$Fe$_{4}$As$_{4}$O$_{2}$}
\newcommand{\rtfao}{RbTb$_{2}$Fe$_{4}$As$_{4}$O$_{2}$}
\newcommand{\rhfao}{RbHo$_{2}$Fe$_{4}$As$_{4}$O$_{2}$}
\newcommand{\rdfao}{RbDy$_{2}$Fe$_{4}$As$_{4}$O$_{2}$}
\newcommand{\rfa}{RbFe$_{2}$As$_{2}$}
\newcommand{\kfa}{KFe$_{2}$As$_{2}$}
\newcommand{\ckfa}{CaKFe$_4$As$_4$}
\newcommand{\cfa}{CaFe$_2$As$_2$}
\newcommand{\gfao}{GdFeAsO}

\newcommand{\btfao}{Ba$_2$Ti$_{2}$Fe$_{2}$As$_{4}$O}
\newcommand{\svfao}{Sr$_2$VFeAsO$_3$}
\newcommand{\kcfaf}{KCa$_{2}$Fe$_{4}$As$_{4}$F$_{2}$}

\begin{document}
\begin{CJK*}{UTF8}{bsmi}
\title{
Study of the rare earth effects on the magnetic fluctuations in RbLn$_2$Fe$_4$As$_4$O$_2$ (Ln = Tb, Dy and Ho) by M\"ossbauer spectroscopy
}
\author{Zhiwei Li}\email{zweili@lzu.edu.cn}
\author{Yang Li}
\affiliation{Institute of Applied Magnetics, Key Lab for Magnetism and Magnetic Materials of the Ministry of Education, Lanzhou University, Lanzhou 730000, P.R. China}
\author{Zhicheng Wang}
\author{Guanghan Cao}
\affiliation{Department of Physics, Zhejiang University, Hangzhou 310027, P.R. China}
\author{Bo Zhang}
\author{Hua Pang}
\author{Fashen Li}
\affiliation{Institute of Applied Magnetics, Key Lab for Magnetism and Magnetic Materials of the Ministry of Education, Lanzhou University, Lanzhou 730000, P.R. China}
\date{\today}

\begin{abstract}
In the current work, we report the investigation of \rlfao\ and \rfa\ by $^{57}$Fe M\"ossbauer spectroscopy. Singlet pattern has been observed for all the samples indicating the absence of static magnetic ordering down to 5.9\,K on the Fe sublattice. The observed intermediate value of the isomer shift confirms the effective self charge transfer effect for the studied superconductors. Debye temperatures of these samples have been determined by the temperature dependence of the isomer shift. Most importantly, we observe different spectral line width broadening behaviors for samples with different rare earth elements, and no line width broadenings for samples Ln=Tb and Ho. These results suggest that the observed magnetic fluctuations at the Fe site might be sensitive to the magnetic behaviors of the rare earth magnetic moments.
\end{abstract}


\maketitle
\end{CJK*}

\section{Introduction}
In the past ten years, many iron-based superconductors (IBS) crystallizing in several structural types have been successfully synthesized \cite{pnas.105.14262,prb.78.060505,cm2009.2189,prl.101.107006,prb.82.180520,jacs.130.3296,ap.59.803,prb.82.104518,jacs.34.12893,jacs.138.3410,jacs.138.7856,cm.29.1805} after the famous discovery of high-temperature superconductivity (HTSC) in LaFeAsO$_{1-x}$F$_x$ \cite{jacs.130.3296}. The key unit responsible for the emergence of HTSC is the so called anti-fluorite-type Fe$_2$X$_2$ (X=As,Se and others) layers.
According to different separating layers, the IBS can be categorized into relatively simple structures, such as (i) 11-type FeSe \cite{pnas.105.14262}, (ii) 111-type (Li/Na)FeAs \cite{prb.78.060505,cm2009.2189}, (iii) 122-type Ba$_{1-x}$K$_x$Fe$_2$As$_2$ \cite{prl.101.107006} and KFe$_2$Se$_2$ \cite{prb.82.180520}, (iV) 1111-type ReFeAsO$_{1-x}$F$_x$ (Re=rare earth elements) \cite{jacs.130.3296,ap.59.803}, and other relatively complex structures with perovskite-like spacer layers, such as \svfao\ \cite{prb.82.104518}, \btfao\ \cite{jacs.34.12893} and more recently discovered examples of 1144-type \ckfa\ \cite{jacs.138.3410} and 12442-type \kcfaf\ \cite{jacs.138.7856} and RbLn$_{2}$Fe$_{4}$As$_{4}$O$_{2}$ (Ln = Sm, Tb, Dy and Ho) \cite{cm.29.1805}.
In the former simple-structure class, SC can be induced by suppressing the spin density wave order with chemical doping or external pressure \cite{ap.59.803}, whereas in the later complex-structure class SC can be induced by internal charge transfer within the materials themselves \cite{jacs.34.12893,jacs.138.3410,jacs.138.7856,cm.29.1805,pc.548.21}

It was found that, subtle changes of the electronic or structural type properties at the local Fe site can lead to very different superconducting properties of the material \cite{ap.59.803}.
Apart from the electronic and structural aspects, possible spin fluctuation mediated pairing mechanism is another mostly discussed topic in the IBS.
M\"ossbauer spectroscopy (MS) is widely accepted to be one of the most sensitive techniques in terms of energy resolution. Since the IBS naturally contain the mostly used M\"ossbauer nuclide, $^{57}$Fe, it is quite natural to use MS to investigate the IBS. In fact, regarding to the study of not only electronic structure but also magnetic properties of this class of materials, many interesting results have already been acquired by MS \cite{SIShylin2015epl, IPresniakov2013jpcm,SLBudko2016prb,ABlachowski2011prb,AOlariu2012njp,TNakamura2012jpsj,MAMcguire2009njp,ZLi2011prb}.

Therefore, it is very interesting to investigate the local electronic and magnetic properties by using MS technique for the recently discovered iron-based materials with internal charge transfer induced HTSC. In our earlier work, we have confirmed the effective charge transfer effect in superconducting \rgfao\ and have found two spectral line width (LW) broadenings that due to magnetic fluctuations of the Fe moments and transferred magnetic fluctuations from the Gd moments \cite{pc.548.21}. Here, in order to study the rare earth dependent behavior of the magnetic fluctuations, we report systematic investigations of the \rlfao\ superconductors with MS. Interestingly, we found that the observed magnetic fluctuations at the Fe site sensitively depend on the magnetic properties at the rare earth site. Then our results invite further studies of these magnetic properties by using other techniques.

\section{Experiments}
Polycrystalline superconductors of \rlfao\ were synthesized by a solid-state reaction method \cite{cm.29.1805}. Phase purity of these samples were checked by powder x-ray diffraction (XRD) measurements with Cu-$K_{\alpha1}$ radiation. As reported earlier, these samples are nearly single phase with Ln$_2$O$_3$ as the only detectable impurity phase whose weight percentages are 3.6\,\%, 3.9\,\% and 2.8\,\% for Ln=Tb, Dy and Ho, respectively \cite{cm.29.1805}. Synthesis and characterization details can be found elsewhere \cite{cm.29.1805}.
\rfa\ single crystal was additionally grown, for comparison of the obtained hyperfine parameters, using self flux method as the same procedure for the growth of CsFe$_2$As$_2$ \cite{prb.87.214509}.

Transmission M\"ossbauer spectra at temperatures between 5.9\,K and 300\,K were taken by using a conventional spectrometer working in constant acceleration mode with a $\gamma$-ray source of 25\,mCi $^{57}$Co(Rh) vibrating at room temperature. The absorbers were prepared with a surface density of $\sim$6\,mg/cm$^2$ natural iron. The drive velocity was calibrated using sodium nitroprusside (SNP) powder and the isomer shifts (IS) quoted in this work are relative to that of the $\alpha$-Fe foil at room temperature. The full line width (LW) at half maximum of the SNP spectrum is 0.244(2)\,mm/s which can be regarded as the resolution of the spectrometer. All the spectra were analyzed with MossWinn 4.0 \cite{mosswinn} programme.

\section{Results and discussion}
\begin{figure}
\centering
\includegraphics[width=0.9\columnwidth,clip=true]{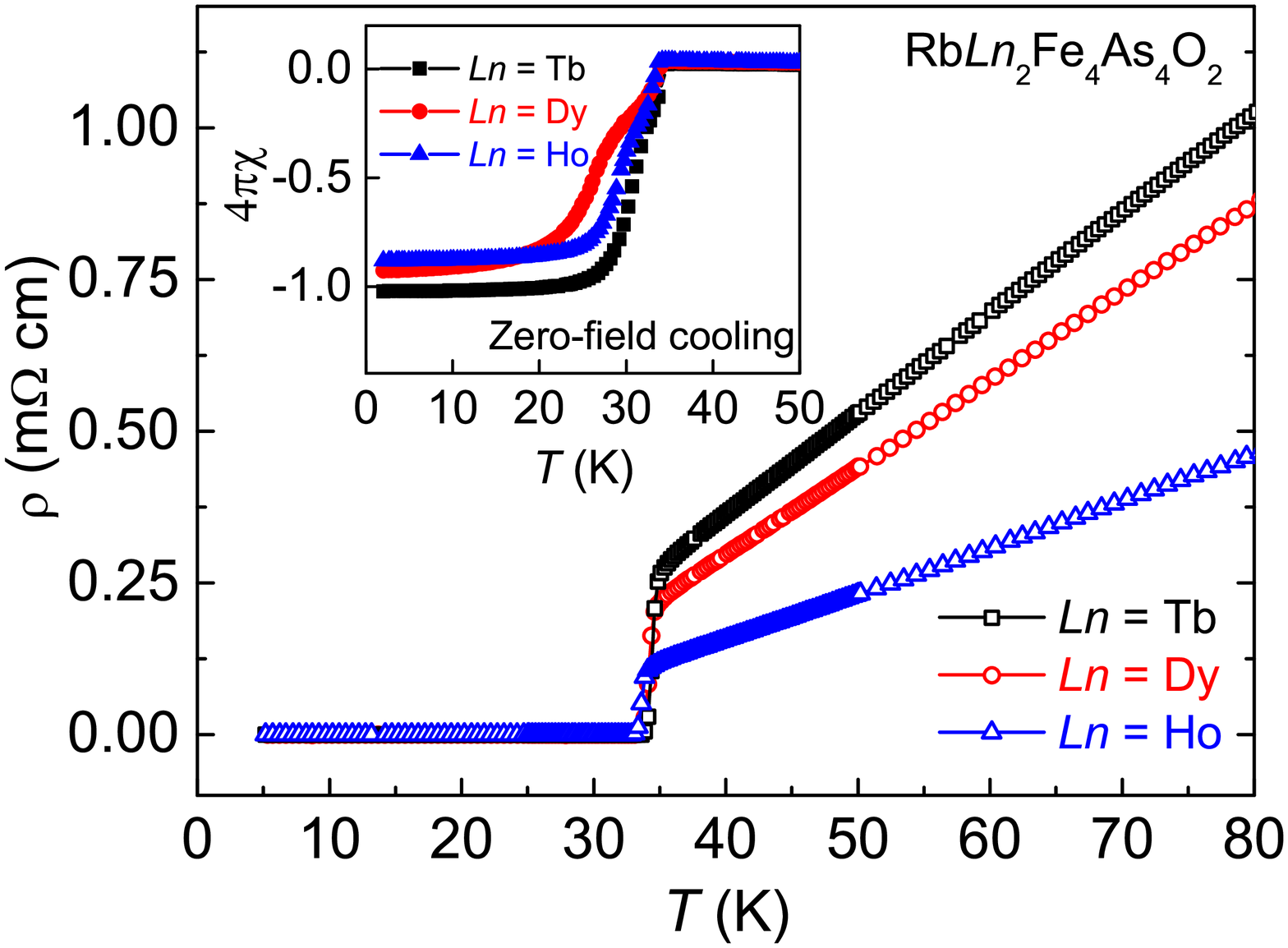}
\caption{
(color online)
Temperature dependence of resistivity for the \rlfao\ superconductors. The superconducting transition temperature were determined to be 34.7\,K, 34.4\,K and 33.8\,K for Ln=Tb, Dy and Ho, respectively. Inset shows the temperature dependence of the $4\pi\chi$ data taken in the ZFC mode with 10\,Oe external magnetic field.
}
\label{Figtc}
\end{figure}

Fig.\ref{Figtc} presents the temperature dependence of the resistivity data, $\rho(T)$, of the as-prepared polycrystalline samples \rlfao.
The superconducting transition temperature were found to be 34.7\,K, 34.4\,K and 33.8\,K for Ln=Tb, Dy and Ho, respectively.
Bulk superconductivity in these samples were confirmed by the strong diamagnetic signal below the superconducting transition temperature in the dc magnetic susceptibility data shown in the inset of Fig.\ref{Figtc}. Detailed superconducting properties of these samples have already been published in our earlier work \cite{cm.29.1805}.

\begin{figure*}
\centering
\includegraphics[width=1.7\columnwidth,clip=true]{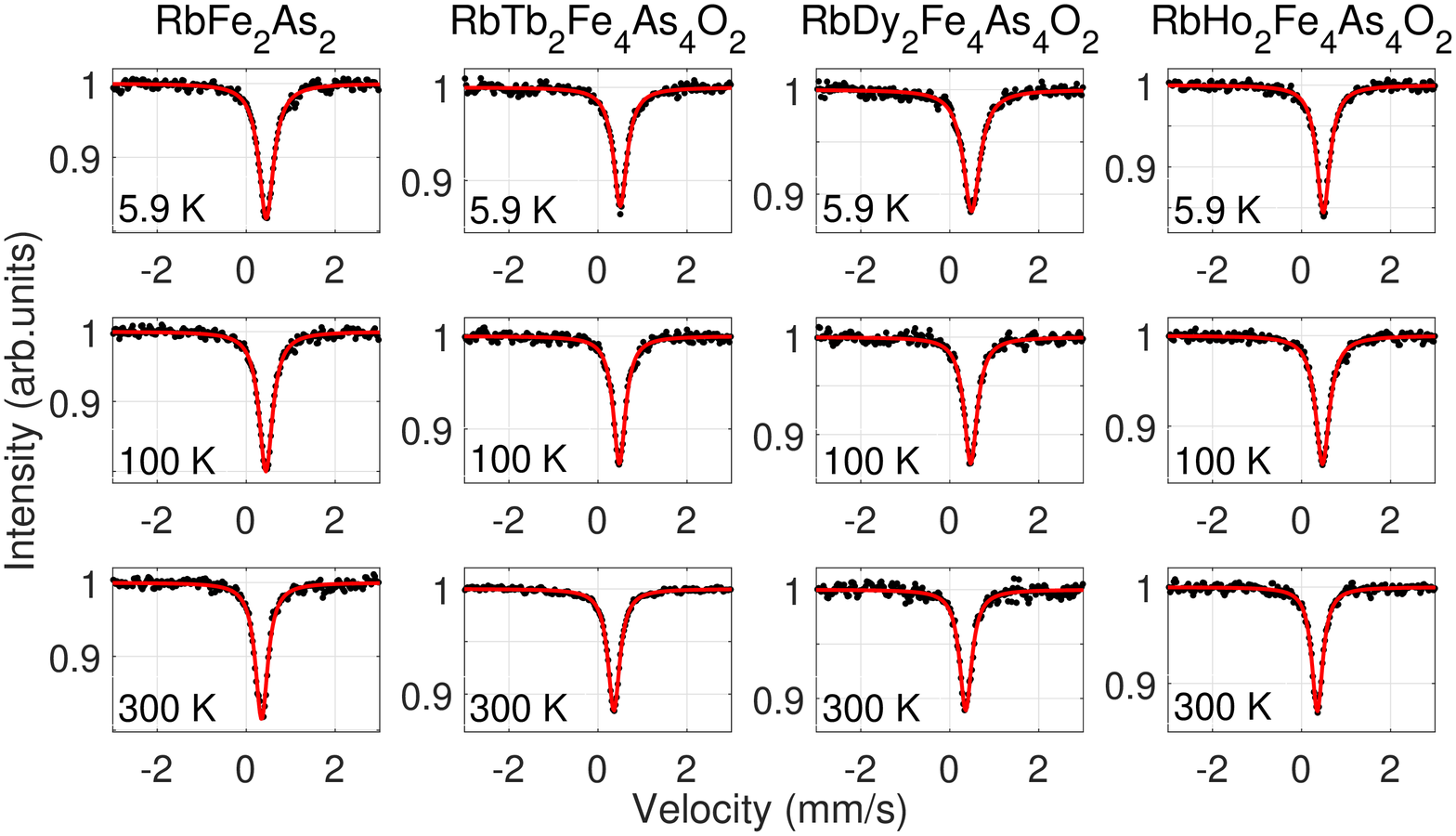}
\caption{
(color online)
$^{57}$Fe M\"ossbauer spectra (black dots) of \rlfao\ and \rfa\ taken at 5.9\,K, 100\,K and 300\,K. The red solid lines are singlet fits to the experimental data.
}
\label{Figmoss}
\end{figure*}

Typical $^{57}$Fe M\"ossbauer spectra of \rlfao\ and \rfa\ samples recorded at temperatures of 5.9\,K, 100\,K and 300\,K are shown in Fig.\ref{Figmoss}. Singlet pattern were observed for all the measured spectra indicating the absence of static magnetic ordering of the Fe sublattice down to the lowest temperature of the current experimental limit of 5.9\,K for all the samples.
In order to get the hyperfine parameters, we fitted our spectra with one singlet profile as described in our earlier work for a similar compound \rgfao\ \cite{pc.548.21}. The good matching between the calculated spectra and the experimental data proves that there is only one Fe species in our samples, being consistent with our XRD analysis that there is only one crystallographic site for Fe atoms and the only detectable impurity phases are Ln$_2$O$_3$ (Ln = Tb, Dy and Ho) \cite{cm.29.1805} which are transparent to our $^{57}$Fe M\"ossbauer measurements since they do not contain Fe element.

\begin{figure}
\centering
\includegraphics[width=0.9\columnwidth,clip=true]{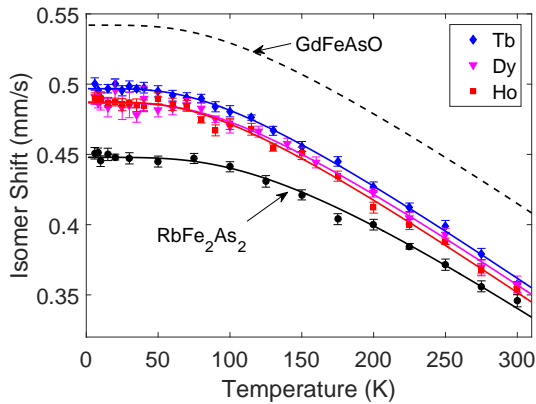}
\caption{
(color online)
Isomer shift as a function of temperature for \rlfao\ derived from the fits as shown in Fig.\ref{Figmoss}. The solid lines are theoretical fits to the Debye model as described in the main text. Data obtained for sample \rfa\ and calculated curve using data taken from Ref.\cite{PWang2010jpcm} for \gfao\ are also shown for comparison.
}
\label{FigIS}
\end{figure}

The ISs obtained from the fits as described above are plotted in Fig.\ref{FigIS} as a function of temperature. Data obtained for sample \rfa\ and calculated curve for \gfao\ using data taken from Ref.\cite{PWang2010jpcm} are also shown for comparison.
At lowest temperature, the ISs of samples \rlfao\ are located in the middle of that for samples \rfa\ and \gfao\ indicating an effective self charge transfer at the local Fe site as found for sample \rgfao\ \cite{pc.548.21} and \ckfa\ \cite{SLBudko2017pm}. With increasing temperature, the ISs gradually decrease following the typical behavior of Debye model. One should note that the ISs of samples \rlfao\ decrease more rapidly than that of sample \rfa\ with increasing temperature, consequently, the ISs are much closer to that of sample \rfa\ at higher temperatures, suggesting different lattice dynamic behaviors of these samples with different characteristic Debye temperatures. On the other hand, the lattice dynamic behaviors of these samples are more similar to that of sample \gfao\ as shown later by their similar Debye temperatures (see table \ref{table}).

In Debye model, the temperature dependence of IS(T) can be expressed by the following equation \cite{Mossbook}
\begin{equation}
\begin{split}
IS(T) = IS(0) - \frac{9}{2}\frac{k_BT}{Mc}(\frac{T}{\Theta_D})^3\int_0^{\Theta_D/T}\frac{x^3dx}{e^x-1}
\end{split}
 \label{eqIS}
\end{equation}
where IS(0) is the temperature independent chemical shift, and the second part is the temperature dependent second order Doppler shift. $k_B$ is the Boltzmann constant, $M$ is the mass of the M\"ossbauer nucleus, $c$ is the speed of light and $\Theta_D$ is the corresponding Debye temperature.
Theoretical fits by using equation (\ref{eqIS}) to the experimental data are shown as solid curves in Fig.\ref{FigIS}. The fitted chemical shift IS(0) and Debye temperature $\Theta_D$ are listed in table \ref{table}. Data taken from Ref.\cite{pc.548.21} for sample \rgfao\ and Ref.\cite{PWang2010jpcm} for \gfao\ are also included for comparison reasons. One can see that the Debye temperature of \rfa\ is much larger than the other samples. It is even larger than that, $\Theta_D=474(20)$\,K, for sample \kfa\ obtained with the same method \cite{SLBudko2017pm}. For the superconducting samples with Ln=Gd, Tb, Dy, and Ho, the smaller Debye temperatures suggest a less stiffer lattice behavior, which should be similar to that of sample \gfao\ since they have much closer Debye temperatures. This is similar to the situation reported for sample \ckfa, which also can be viewed as being constructed from the alternating slabs of \cfa\ and \kfa, whose Debye temperature is also much smaller than that of \kfa\ and much closer to the value of \cfa\ \cite{SLBudko2017pm,jpcm.23.255701}.

\begin{table}[ht]
\centering
\caption{Chemical shift IS(0) and Debye temperature $\Theta_D$ of samples \rlfao\ and \rfa\ obtained by fitting eqution (\ref{eqIS}) to the data in Fig.\ref{FigIS}. Data taken from Ref.\cite{pc.548.21} for sample \rgfao\ and Ref.\cite{PWang2010jpcm} for \gfao\ were also included.}
\label{table} {
\begin{tabular}{c c c}
\hline \hline
Sample & IS(0) (mm/s)  &    $\Theta_D$ (K)      \\
\hline
\rfa	& 0.448(2)	& 526(28)	   \\
\rgfao	& 0.497(4)	& 415(47)	    \\
\rtfao	& 0.497(1)	& 366(15)	    \\
\rdfao	& 0.487(2)	& 395(25)	    \\
\rhfao	& 0.487(2)	& 363(20)	    \\
\gfao	& 0.540(2)	& 409(4)	   \\
\hline \hline
\end{tabular}}
\end{table}

\begin{figure}
\centering
\includegraphics[width=0.9\columnwidth,clip=true]{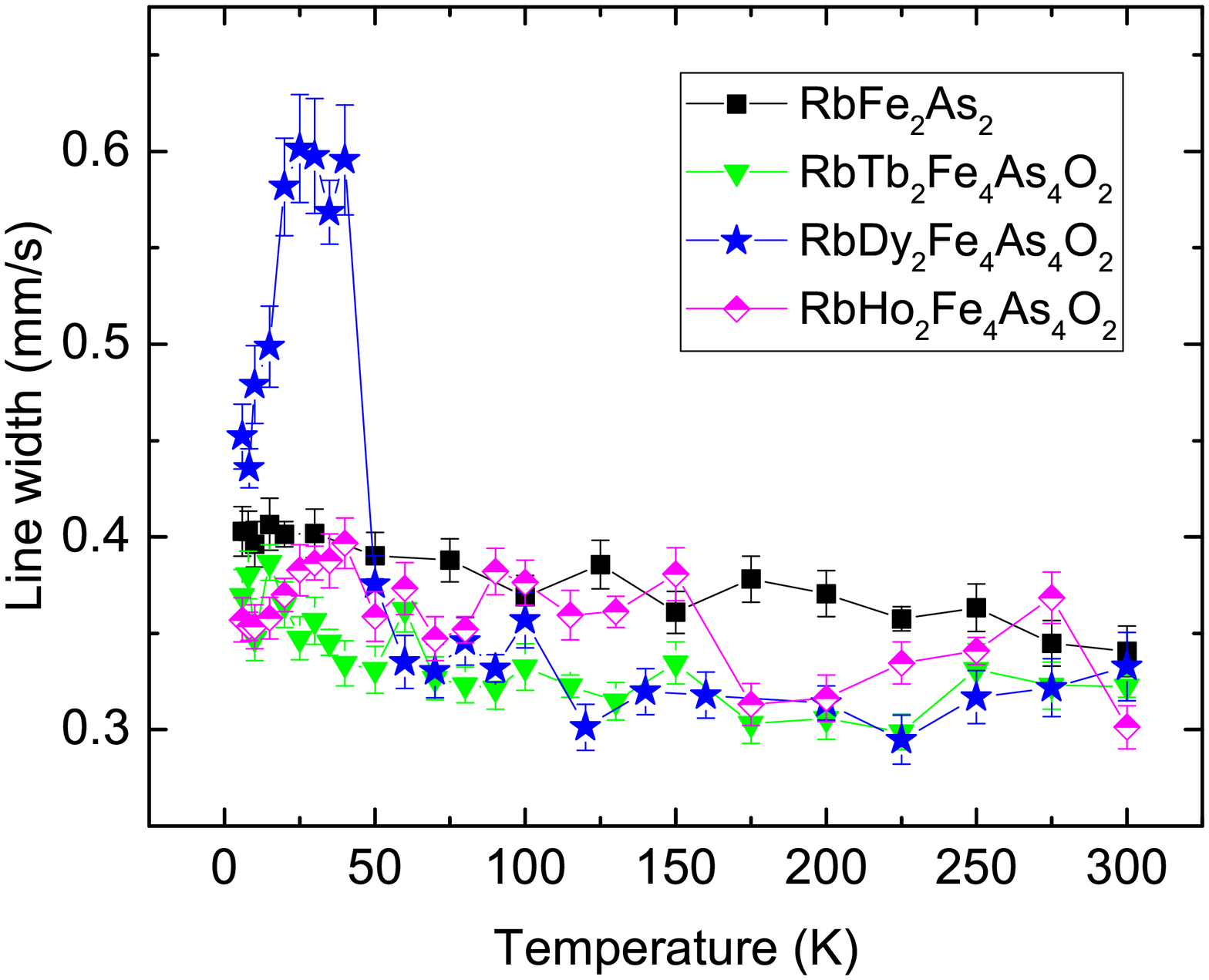}
\caption{
(color online)
Temperature dependence of the spectral LW of \rlfao\ and \rfa.
}
\label{FigLW}
\end{figure}

In Fig.\ref{FigLW}, we plotted the spectral LW of \rlfao\ and \rfa, determined from the fits shown in Fig.\ref{Figmoss}, as a function of temperature.
 A featureless temperature dependent behavior can be observed for sample \rfa\ and samples with Ln=Tb, Ho. On the other hand, a spectral LW broadening can be seen below about 50\,K for sample with Ln=Dy. This is similar with our earlier report for sample \rgfao\ where two spectral LW broadenings have been observed with decreasing temperature below about 100\,K and 15\,K corresponding to magnetic fluctuations from the Fe spins and transferred magnetic fluctuations from the Gd spins \cite{pc.548.21}, respectively.
 With further decreasing temperature, the LW broadening for sample Ln=Dy has been suppressed within the superconducting ground state due to the competition between magnetism and SC.
 Different with sample \rgfao, no second LW broadening has been observed till the lowest measured temperature of 5.9\,K in this study, which suggests that no transferred magnetic field can be observed at the Fe site for sample with Ln=Dy. This is rather confusing since that the magnetic ordering temperature for sample Ln=Dy ($\sim$8\,K) is higher than that for sample Ln=Gd ($\sim$3\,K, see the magnetic measurements in Ref.\cite{cm.29.1805,pc.548.21}) and the magnetic moment is usually larger for Ln=Dy. From the discussion in our earlier work \cite{pc.548.21}, we believe that the transferred magnetic fluctuations are most likely due to the Gd moments that order in a canted antiferromagnetic fashion below a lower temperature. Therefore, our results suggest that the magnetic ordering of the Dy moments observed for sample Ln=Dy is likely to be collinear rather than canted for sample Ln=Gd. Below the ordering temperature, the magnetic field at the Fe site transferred from the Dy magnetic moments on the two sublattices shall cancel out. Thus, the absence of the second LW broadening can be understood.
 From the earlier published work \cite{cm.29.1805}, no magnetic ordering of the rare earth magnetic moments have been observed for sample Ln=Tb and Ho down to $\sim$2\,K. Then, the absence of spectral LW broadening for sample Ln=Tb and Ho can be understand naturally.
 The above results suggest that the observed LW broadenings, either due to the magnetic fluctuations of the Fe spins or due to the transferred fluctuations from the rare earth spins, might be sensitive to the magnetic behaviors of the rare earth moments.
 This is rather surprising since no obvious dependence of the superconducting properties on the rare earth magnetism were observed in the past. Therefore, finding out how and when do the rare earth magnetic moments order is very important for a better understanding of these phenomena.

\section{Summary}

In summary, we have investigated \rlfao\ and \rfa\ by $^{57}$Fe M\"ossbauer spectroscopy. The singlet pattern of these spectra indicate that all these samples do not exhibit static magnetic ordering down to 5.9\,K on the Fe sublattice. The effective self charge transfer effect for our \rlfao\ samples have been confirmed by the observed intermediate value of the isomer shift. Debye temperatures of these samples have been determined by the temperature dependence of the isomer shift. Importantly, the different spectral LW broadening behaviors observed for sample Ln=Dy and Gd, and the absence of the LW broadening for sample Ln=Tb and Ho suggest that the observed magnetic fluctuations at the Fe site might be sensitive to the magnetic behaviors of the rare earth magnetic moments. Therefore, future investigations using other techniques to study these effects and the possible interplay between magnetism and superconductivity in these materials will be very interesting.

\section{Acknowledgements}

This work was supported by the National Natural Science Foundations of China (No. 11704167), the Fundamental Research Funds for the Central Universities (No. lzujbky-2017-31) and the National Key Research and Development Program of China (No. 2016YFA0300202).

\bibliography{rrefao}

\end{document}